\documentclass [12pt,a4paper,draft]{article}

\usepackage{times}

\DeclareFontFamily{OT1}{times}{}
\DeclareFontShape {OT1}{times}{m }{n }{ <-> ptmr }{}
\DeclareFontShape {OT1}{times}{bx}{n }{ <-> ptmb }{}
\DeclareFontShape {OT1}{times}{m }{it}{ <-> ptmri}{}
\DeclareFontShape {OT1}{times}{bx}{it}{ <-> ptmbi}{}
\usepackage{amsmath}
\usepackage{amsfonts}
\usepackage{amssymb}
\usepackage{latexsym}
\newcommand{\cl}{C \kern -0.1em \ell} 

\newcommand{\CON}{\overline}          

\newcommand{\vect}{\wedge}            

\newcommand{\SCA}{\langle}            
\newcommand{\LAR}{\rangle}            
\newcommand{\BRA}{\langle\kern -.2em\langle} 
\newcommand{\KET}{\rangle\kern -.2em\rangle} 

\newcommand{\REA}{\operatorname{Re}}  
\newcommand{\IMA}{\operatorname{Im}}  

\newcommand{\Q}{[\hspace{1.mm}]} 
\newcommand{\A}{(\hspace{.5mm})} 

\newcommand{\REV}{\sim}               

\newcommand{\Oh}{\tfrac{1}{2}}    
\newcommand{\Th}{\tfrac{3}{2}}    
\begin{document}

\title{\bf{Lanczos's equation as a way out\\ of the spin 3/2 crisis?} \\ 
~\\ }

\author{{\bf André Gsponer and Jean-Pierre Hurni}\\
\emph{Independent Scientific Research Institute}\\ 
\emph{ Box 30, CH-1211 Geneva-12, Switzerland}\\
e-mail: isri@vtx.ch\\}

\date{ISRI-02-08 ~~ \today }               

\maketitle

\begin{abstract}

\normalsize
It is shown (1) that Lanczos's quaternionic formulation of Dirac's equation does not lead to a solution of the problems that plague the standard spin 3/2 theory based on the Rarita-Schwinger equation, but (2) that the four-component solutions to the quaternionic generalization of Dirac's equation proposed by Lanczos in 1929 may provide a consistent theory for spin 3/2 particles, although at the cost of giving up the postulate that there should be a one-to-one correspondence between arbitrary-high-spin unitary representations of the inhomogeneous Lorentz group and elementary particles.

\end{abstract}	

~\\
~\\

\begin{center}


\end{center}

\newpage

\section{Introduction}

In the years before 1928 the main experimental difficulty motivating the search for a relativistic wave equation was the problem of spin effects in atomic spectra that could not be explained by either the non-relativistic Schr\"odinger equation or the relativistic Klein-Gordon equation.  Thus, by discovering the equation that bears his name, Dirac solved the first ``spin 1/2 crisis.'' However, despite the remarkable success of Dirac's equation in atomic physics and quantum electrodynamics, many problems related to the intrinsic spin of elementary particles are still far from being solved today.\footnote{We use the definition that a ``truly elementary'' particle is a pointlike object such as an electron, and an ``elementary'' particle an object such as a hadron whose constituents (the quarks) cannot be isolated as free pointlike objects. Elementary particles may have excited states (resonances) but truly elementary particles should not.} 

For instance, if the spin of the photon and of the intermediate vectors bosons can be considered as understood, and if the spin and magnetic moment of the muon seem to be fully described by Dirac's equation, there is no comparable understanding of the proton spin, and more generally neither of the spin or magnetic moment of any hadron.


For example, the intrinsic magnetic moment of a Dirac particle is directly proportional to its electric charge.  The neutron magnetic moment should therefore be zero (which is experimentally not the case) so that its non-zero magnetic moment is qualified as ``anomalous.''  Successful first order predictions of the neutron magnetic moment, and of many other properties of hadrons, are given by the quark model. Unfortunately this model suffers from a number of internal inconsistencies (such as the unsolved problem of quark confinement) to  which a considerable difficulty was recently added, the so-called ``nucleon spin (or spin~1/2) crisis:''  the measured quark spin contribution is only a small fraction of the nucleon spin, in contradiction with the naive quark model!  

But there is a further major problem with spin, which we may call the ``spin~3/2  crisis.'' It consists of the fact that there is at present no consistent theory of elementary particles with spin equal to or higher than $\Th$.  For instance, while it is possible to build many reasonable classical wave equations for \emph{free}  particles of spin $\Th$ or higher, all equations proposed until now lead to problems when a non-vanishing external field is introduced.  Mathematically, this is not in contradiction with Wigner's proposal that there should be a one-to-one correspondence between arbitrary-high-spin unitary representations of the inhomogeneous Lorentz group and elementary particles \cite{WIGNE1939-}.  Indeed, this classification applies only to \emph{free}, i.e., noninteracting particles \cite[p.213]{BARGM1948-}.  The problem is that interactions are an absolute necessity of a realistic physical theory.

Therefore, soon after Dirac proposed a system of equations applicable to particles of arbitrary spin \cite{DIRAC1936-}, Fierz and Pauli showed that because these equations imply too many independent components, supplementary conditions are necessary when the particles are coupled to an external electromagnetic field \cite{FIERZ1939-}.  Moreover, when these equations are quantized, the corresponding field theories are inconsistent.  For example, in the spin $\Th$ case, they yield non-positive-definite fermion anticommutators, a problem first discovered by Johnson and Sudarshan  \cite{JOHNS1961-}. Looking more closely at the origin of these inconsistencies, Velo and Zwanziger discovered that they arise already at the classical level in the form of noncausal modes of propagation, the so-called ``Velo-Zwanziger phenomenon'' \cite{VELO-1969A}.  Finally, when high-spin equations are used to calculate scattering amplitudes, one obtains diverging results that violate unitarity in the high-energy limit.  To get well-behaved amplitudes it is necessary to impose ad-hoc conditions such as adding an explicit magnetic dipole moment to the minimal Lagrangian \cite{FERRA1992-}.

Consequently, it has become widely accepted that there should, most probably, not exist any truly fundamental particle with spin higher than 1 (see, e.g.\ \cite[p.11]{VELO-1978-}) which is a limiting case since spin 1 particles also exhibits inconsistencies when non-minimally coupled to an external field \cite{VELO-1969B}.\footnote{The problems with high-spin wave equations and their associated field theories have been extensively reviewed in a remarkable, but little cited, series of lectures on Invariant Wave Equations given at the \emph{International School of Mathematical Physics ``Ettore Majorana''} that took place from June 27 to July 9, 1977, in Erice, Italy \cite{VELO-1978-}.  This book is the most recent major publication on spin~$\Th$ included in our list of references.  This is because since about that time there has been no truly significant advance towards the resolution of the fundamental problems related to high-spin equations.  In fact, most recent publications (e.g., in supersymmetry theory, or high-energy particle phenomenology) use the Rarita-Schwinger and other high-spin equations as if these fundamental problems did not exist!  Nevertheless, some progress has been made in applications which do not crucially depend on these problems, for example by the ``CALKUL collaboration'' in the evaluation of multi-particle amplitudes by the so-called helicity method \cite{NOVAE1992-}.}

However, we know from experiment that elementary spin $\Th$ particles exist: the members of the baryon decuplet, i.e., the $\Delta, \Sigma, \Xi,$ and $\Omega$ particles.  While they are short-lived, and ``non-truly elementary'' in the sense that they are composed of confined quarks, these particles are not composite objects comparable to atoms or nuclei which are made of electrons and nucleons that can freely exist.  Their existence as elementary entities which are more than just exited states of other particles has therefore to be explained.

Moreover, there are a number of theoretical reasons for thinking that elementary rather than composite spin $\Th$ particles have to exist.  For instance, in supersymmetric theories, spin $\Th$ particles have to exist as fermionic partners of spin 2 bosons \cite{MORIT1985-}; in twistor theory, massless fields of helicity  $\Th$ arise in connection with the Einstein equations of gravitation \cite{PENRO1992-}; and in the quaternionic generalization of Dirac's equation proposed by Lanczos in 1929, spin $\Th$ states appear on the same footing as those of spin 0, $\Oh$, and 1, see \cite{LANCZ1929-,GSPON1994-,GSPON1998-}.  Thus, there are many compelling reasons for trying to find a working theory for spin $\Th$ particles and fields.

In this paper, we leave aside the general question of particles of spin 2 and higher, and investigate whether a quaternionic\footnote{Biquaternions enable special relativity, Maxwell's and Dirac's equations, Lanczos's equation, as well as their implications in classical and quantum theory, to be expressed in particularly concise but explicit forms. Since the algebra of biquaternions $\mathbb{B}$ is four-dimensional over the complex numbers, it has the minimum number of components necessary to deal with spin $\Th$ particles.}  formulation of elementary field theory could help finding a solution to the ``spin 3/2 crisis.''  More precisely,  we inquire

\begin{itemize}
\item whether the use of Lanczos's quaternionic formulation of Dirac's equation (i.e., the \emph{Dirac-Lanczos equation}) could avoid the ``Velo-Zwanziger phenomenon'' that plagues the standard spin $\Th$ theory based on the Dirac-Rarita-Schwinger equation \cite{RARIT1941-};\footnote{As emphasized by Wightman, see \cite[p.3]{VELO-1978-}, the Rarita-Schwinger formalism provides, with some change in notation, Dirac's 1936 proposal for a spin $\Th$ particle \cite{DIRAC1936-}.} and

\item  whether the spin $\Th$ solutions to the quaternionic generalization of Dirac's equation proposed by Lanczos in 1929 (i.e.,  \emph{Lanczos's fundamental equation}, which incorporates isospin and provides a common framework for elementary fields of spin 0, $\Oh$, 1, and $\Th$) could be acceptable for describing the experimentally observed spin $\Th$ particles.
\end{itemize}

The outline of the paper is as follows.  In Section~2 we give a simple argument showing that problems are due to arise when dealing with particles and fields of spin greater than 1; in Section~3 we summarize the features of Lanczos's generalization of Dirac's equation that are relevant to the present paper; in Section~4 we investigate whether the use of the \emph{Dirac-Lanczos equation} could solve some problems related to the Rarita-Schwinger formalism; in Section~5 we study the implications of the postulate that the general four-component solutions of \emph{Lanczos's fundamental equation} could correspond to elementary spin $\Th$ particles; and, in Section~6, we conclude and speculate on the idea that the observed spin $\Th$ particles could shed some new light on the nature of fundamental particles, a ``ten-fold way'' perspective which has not be fully exploited until now.

\section{Origin of the spin 3/2 problem}

There are several problems with high-spin particle theories, and it is not clear whether these problems have a common origin. However, as with many other outstanding problems in physics, the ``spin 3/2 crisis'' is basically a clash between special relativity and quantum theory.  Indeed, the existence of potential problems can be seen at the elementary level already, i.e., without making any reference to wave equations or field theory.  This is because there are latent conflicts between the orthogonal scalar product $\SCA\CON{X}Y\LAR$ of special relativity and the unitary scalar product $\SCA{X}^+Y\LAR$ of quantum theory.\footnote{Our notations are defined in the Appendix which is based on reference \cite{GSPON2001-}.} 

Consider two fields $X$ and $Y$ of spin $s$ and the general Lorentz transformation $\mathcal{L}_s\A$ operating of these fields.  This general transformation can decomposed into the product of a rotation $\mathcal{R}_s\A$ and a boost $\mathcal{B}_s\A$. It is then a matter of simple algebra to verify that for both $s=\Oh$ and $s=1$ we have the relations
$$
    \SCA \CON{                     X}               Y  \LAR =
    \SCA \CON{\mathcal{L}}_s(\CON{X})\mathcal{L}_s(Y) \LAR =
    \SCA \CON{\mathcal{R}}_s(\CON{X})\mathcal{R}_s(Y) \LAR =
    \SCA \CON{\mathcal{B}}_s(\CON{X})\mathcal{B}_s(Y) \LAR   \eqno(1)
$$
for the special relativity scalar product, while for the quantum scalar product we have only
$$
    \SCA                 X^+               Y  \LAR =
    \SCA \mathcal{R}_s^+(X^+)\mathcal{R}_s(Y) \LAR   ~~.  \eqno(2)
$$
because unitary implies that boosts are excluded.

  However, if $X$ and $Y$ are two four-component spin $\Th$ spinors, the special relativity scalar product $\SCA\CON{X}Y\LAR$ is \emph{not} conserved and one has only
$$
    \SCA                     X^+                   Y  \LAR =
    \SCA \mathcal{R}_{3/2}^+(X^+)\mathcal{R}_{3/2}(Y) \LAR   ~~.      \eqno(3)
$$
There is therefore a striking difference between the spin $\Oh$ and 1 fields (to which one can trivially add the spin 0 field) and the fields of spin $s \geq \Th$.  This difference is such that while the conservation of the quantum scalar product under spatial rotations implies that spin is a ``good quantum number'' --- meaning that it is possible to construct unitary representations of the Poincaré group for free particles of arbitrary-high spin --- there are problems as soon as these fields are coupled to spin $\Oh$ and 1 fields, i.e., to elementary fermions and to the electroweak field.  For instance, it is in general not possible to construct multilinear quantities mixing spin $\Th$ fields with  spin $\Oh$ or 1 fields that are covariant under both the quantum and the special relativity scalar products.

Because of this fundamental difference, it is of interest to look somewhat closer at the geometry of spin $\Th$ rotations, a topic on which  text-books do not say very much. The reason is obvious: while a spin 1 rotation is simply a familiar rotation in ordinary space, a spin $\Oh$ rotation is already something less intuitive since it implies that a rotation angle of $4\pi$ (instead of $2\pi$) is required to fully turn around a spin $\Oh$ object such as an electron or a neutron.  Nevertheless, the rotation of a spin $\Oh$ field $\phi$ by an angle $\theta$ around an axis defined by a unit vector $\vec{a}$ has the very simple form
$$
 {\phi}' = \mathcal{R}_{1/2}(\phi)   \hskip 1in
           \mathcal{R}_{1/2}\A = \exp(\Oh\theta\vec{a})\Q   \eqno(4)
$$
where $\phi$ is a singular biquaternion so that it corresponds to only four real numbers which are isomorphic to the two complex components of a spin $\Oh$ spinor.\footnote{A biquaternion $S$ is singular, or null, if $S\CON{S}=0$. Conversely, $Q$ is regular, or invertible, if $Q\CON{Q}\neq 0$.} Similarly, the rotation of a spin 1 field, whose three complex components correspond to a complex vector  $\vec{\eta}$ in biquaternions, is given by the well-known Olinde-Rodrigues formula
$$
 \vec{\eta'} = \mathcal{R}_{1}(\vec{\eta})   \hskip 1in
               \mathcal{R}_{1}\A = \exp( \Oh\theta\vec{a}) ~ \Q ~
               \exp(-\Oh\theta\vec{a})   ~~.                \eqno(5)
$$

In comparison, even with quaternions, a spin $\Th$ rotation is much more complicated.  In fact, it cannot be expressed in a form such as $(4)$ or $(5)$ where the argument is simply multiplied from either side by a quaternion.  To write down such a rotation, and to compare it to the spin $\Oh$ and 1 cases, one has to express it as an exponential in the Lie generators of the corresponding representations of the rotation group.  To do that in a systematic way (and to find the corresponding rotation eigenstates) one can use the so-called ``Peirce decomposition'' theorem which enables to write any biquaternion $\xi \in \mathbb{B}$ as a linear combination of two idempotents ($\sigma$ and $\CON{\sigma}$) and two nilpotents ($\tau\sigma$ and $\tau\CON{\sigma}$),\footnote{An idempotent $\sigma = \sigma^2$ has the property that  $\sigma\overline\sigma = 0$. For definitiveness,  we  take  $\sigma=\Oh(1+i\vec\nu)$  where $\vec\nu$ is a unit vector.  Then $\vec\tau \sigma$ is a nilpotent when $\vec\tau$ is a unit vector perpendicular to $\vec\nu$, i.e., $(\vec\tau \sigma)^2=0$. To simplify the notation we will not always put an arrow over the unit vectors $\nu$ and $\tau$.} i.e.,
$
\xi=x_1\sigma +x_2\tau\sigma +x_3\CON{\sigma} +x_4\tau\CON{\sigma} 
$
where $x_n \in \mathbb{C}$.  Then, if the $a_n$ are the three projections  $\vec{a}$ on the orthogonal triad $\tau\nu$, $\tau$, and $\nu$, any general spin $s$ rotation around $\vec{a}$ can be written
$$
\xi' = \mathcal{R}_{s}(\xi)   \hskip 1in
       \mathcal{R}_{s}\A = \exp\bigl(-i\theta\sum_1^3 a_n J_n\A\bigr)  \eqno(6)
$$ 
where the $J_n\A$, given in Table 1, are the quantum mechanical spin generators corresponding to spin $\Oh$, 1, or $\Th$ rotations.\footnote{With the substitution $\theta \rightarrow i\rho$, which replaces the angle $\alpha$ by the rapidity $i\rho$, one obtains the general expression for a Lorentz boost.}  If expression $(6)$ is calculated for the spin $\Oh$ and 1 cases one falls back on $(4)$ and $(5)$. In the spin $\Th$ case, however, one cannot find a simple expression and it is better to remain with the general form~$(6)$.

Coming back to the problem of understanding the origin of the ``spin 3/2 crisis,'' we see that while it is easy to build multilinear covariant quantities by simply multiplying spin $\Oh$ and spin 1 fields which transform as $(4)$ or $(5)$, this is not possible with spin $\Th$ fields transforming according to $(6)$.\footnote{The same problem obviously arises when using other formalisms.  However, with matrices for example, it is not a trivial result that spin $\Oh$ and 1 transformations can be written as $(4)$ and $(5)$--- which immediately show how to meaningfully combine spin $\Oh$ and spin 1 fields.  Examples: For rotations, the bilinear covariants $\phi_1\CON\phi_2$ and $\vec\eta_1 \vec\eta_2$ are spin 1 fields, $\vec\eta \phi$ is a spin $\Oh$ field, the scalar product $\SCA \vec\eta_1 \vec\eta_2 \LAR$ is an invariant, etc.} In fact, in order to build covariant quantities involving spin $\Th$ fields one has to decompose them into their components, and mix them with individual components of the spin $\Oh$ or spin 1 fields.  This is of course the reason why the spinor calculus has been created, and the explanation of its power in writing high-spin equations in a very concise manner \cite{FIERZ1939-}.\footnote{The conciseness of the spinor calculus is somewhat misleading when it comes to write down explicitly the corresponding equations, which have a form that is not much different from the matrix expressions given by Dirac in \cite{DIRAC1936-}.  To see how this translation is made in the case of spin $\Th$, see \cite{BELIN1953-}.} But this is also a reason why one can be suspicious about the conventional approach --- an approach that has been taken on the assumption that there should exist infinite series of elementary particles with arbitrary-high spin, while the elementary particles that have been found so far all have $s \leq \Th$.

\begin{table}
\hspace{-2mm}
\begin{tabular}{|c|c|c|c|c|}
\hline
           &                          &                     &                   &                  \\
{\bf Spin} &     {\bf 1/2$_+$ }       &    {\bf 1/2$_-$}    &     {\bf 1}       & {\bf 3/2}        \\
           &                          &                     &                   &                  \\
\hline
\hline 
$J_1\A$ & $\Oh i \tau\nu\Q\sigma$ & $\Oh i \tau\nu\Q\CON{\sigma}$ & $\Oh i (\tau\nu\Q-\Q\tau\nu)$ & $-\Oh \tau( ~~\Q \tau + \sqrt{3} \nu\Q \nu + \nu\Q \nu\tau ) $\\
\hline
$J_2\A$ & $\Oh i  ~~\tau\Q\sigma$ & $\Oh i  ~~\tau\Q\CON{\sigma}$ & $\Oh i (~~\tau\Q-\Q\tau~~)$ & $-\Oh \tau(\nu\Q \tau + \sqrt{3}  ~~\Q \nu -  ~~\Q \nu\tau ) $  \\
\hline
$J_3\A$ & $\Oh i   ~~\nu\Q\sigma$ & $\Oh i   ~~\nu\Q\CON{\sigma}$ & $\Oh i (~~\nu \Q-\Q\nu~~)$ & $~~~\Oh \,i (\nu\Q~~ + ~~~ 2 ~~\Q \nu ) $  \\
\hline
\hline
+3/2  &   &   &   & $\sqrt{2}      \sigma     $ \\
\hline
+1    &   &   & $\sqrt{2}      \sigma \tau$  &   \\
\hline
+1/2  & $\sqrt{2}      \sigma     $ & $\sqrt{2}     {\sigma}\tau$ &   & $\sqrt{2}\CON{\sigma}\tau$ \\
\hline
~~0   &   &   & $ \nu$                      &   \\
\hline
--1/2  & $\sqrt{2}\CON{\sigma}\tau$ & $\sqrt{2}\CON{\sigma}    $ &   & $\sqrt{2}      \sigma \tau$ \\
\hline
--1    &   &   & $\sqrt{2} \CON{\sigma}\tau$ &   \\
\hline
--3/2  &   &   &   & $\sqrt{2}\CON{\sigma}    $ \\
\hline

\end{tabular}
\caption{\emph{Spin angular momentum operators $J_n$, and normalized $J_3$-eigenstates, for fields of spin $\Oh$, $1$, and $\Th$. $\sigma$ is the idempotent $\Oh(1+i\nu)$ where $\nu$ is a unit vector. $\tau$ is a unit vector perpendicular to $\nu$. The two spin $\Oh$ columns correspond, conventionally, to isospin ``up''} (1/2$_+$) \emph{and ``down''} (1/2$_-$).}  
\end{table}

\section{Lanczos's equation}

Cornelius Lanczos's PhD dissertation of 1919 was a very ambitious field theory in which electrons and protons were singular solutions of Maxwell's equations written in biquaternion form.  Ten years later, while working in Berlin as Einstein's assistant, Lanczos immediately saw how to fit Dirac's equation into a more general framework that would largely escape the attention of his contemporaries \cite{LANCZ1929-, GSPON1994-, GSPON1998-}. Today, with our hindsight and all the experimental knowledge that has been accumulated, it is easy to reconstitute Lanczos's bold step.

In the standard spinor or two-component formalisms,  Dirac's  equation can be symbolically written as  
$$ 
  \overline\partial  L = m R ~~,  \hskip 1 in     \partial  R = m L   \eqno(7) 
$$
where $L$ and $R$ are the left- and right-handed parts of Dirac's four-component bispinor, and $\partial$ the spinor or Pauli-matrix four-gradient. As is well known, equation $(7)$ can be rewritten in many equivalent forms, using in particular Clifford algebras.  With biquaternions, one can use the ``Peirce decomposition'' and the trivial isomorphisms
$$
\begin{pmatrix}
   c_1     \\
   c_2     \\
\end{pmatrix}
\sim 
\begin{pmatrix}
  c_1  &  0  \\
  c_2  &  0  \\ 
\end{pmatrix} 
\sim (c_1 + c_2\tau) \sigma  = S_1   ~~.     \eqno(8')
$$
or
$$
\begin{pmatrix}
   c_3     \\
   c_4     \\
\end{pmatrix}
\sim
\begin{pmatrix}
 0 & c_3  \\
 0 & c_4  \\ 
\end{pmatrix} 
\sim (c_3\tau + c_4) \CON{\sigma}  = S_2    ~~.     \eqno(8'')
$$
to replace the two-component spinors $L$ and $R$ by the singular quaternions $L_1$ and $R_1$, or $L_2$ and $R_2$. Thus, Dirac's equation in two-component form can be written as
$$
  \overline\nabla L_1 = m R_1 ~~, \hskip 1 in
           \nabla R_1 = m L_1 ~~,                     \eqno(9')
$$
or equivalently as
$$
  \overline\nabla L_2 = m R_2 ~~, \hskip 1 in
           \nabla R_2 = m L_2 ~~.                     \eqno(9'')
$$
 This leads to the fecund idea that the two possibilities $(9')$ and $(9'')$ could in fact be combined in such a way that Dirac's equation $(7)$ would just be a special case of a more fundamental biquaternion equation.  Indeed, this is what Lanczos postulated in 1929 when he wrote his generalized ``Dirac'' equation \cite{LANCZ1929-}
$$
  \overline\nabla A = m B ~~, \hskip 1 in    \nabla B = m A ~~.     \eqno(10)
$$
Here $\nabla = \partial_{it}+\partial_{\vec x}$  is the quaternion four-gradient operator,  assumed to  transform  as  a four-vector, i.e., $\nabla^\prime=\mathcal{L}\nabla\mathcal{L}^+$.  Equation $(10)$ is \emph{Lanczos's fundamental equation} from which two Dirac equations can be derived by projecting out either the $S_1$ or $S_2$ parts of the superpositions $A=L_1+L_2$ and $B=R_1+R_2$, i.e., $L_1=A\sigma$, etc., to isolate the corresponding two-component spinors.

    Lanczos's equation has many remarkable properties and a rich particle content. First, by making appropriate superpositions of the $A$ and $B$ fields it describes an isospin doublets of spin $\Oh$ particles
$$
 \Psi_+ =           A \sigma + B^* \overline\sigma  ~~, \hskip 1 in 
 \Psi_- = (A \overline\sigma - B^*          \sigma) i\tau \nu ~~.\eqno(11)
$$
Both of these linearly independent superpositions satisfy the Klein-Gordon equation, $\nabla\overline\nabla \Psi = m^2 \Psi$, as well as the  {\it Dirac-Lanczos equation}   
$$
    \overline\nabla  \Psi = m \Psi^* i  \nu ~~.   \eqno(12)
$$
This equation is strictly equivalent to Dirac's equation for a four-component bispinor.\footnote{The difference with Dirac's equation $i\gamma^\mu\partial_\mu\psi = m\psi$ is that while the Dirac bispinor $\psi$ is composed of a pair of spinors $\{L,R\}$ which transforms under the representation $\mathcal{D}(\Oh,0)\oplus\mathcal{D}(0,\Oh)$ of the Lorentz group, the Dirac-Lanczos bispinor $\Psi$ is composed of a pair of spinors $\{A\sigma,B^*\overline\sigma\}$ which both transform under the representation $\mathcal{D}(\Oh,0)$.} The spin $\Oh$ character of the superpositions $(11)$ is enforced by the postmultiplication by the constant idempotent $\sigma$, and that of equation $(12)$ by the vector $\nu$, which imply that under a Lorentz transformation one necessarily has $\Psi^\prime=\mathcal{L}\Psi$.\footnote{The invariant unit vectors $\nu$ and $\tau$ define two arbitrary mutually orthogonal directions.  Table 1 shows that $\nu$ is associated with $J_3$, so that $\nu$ is the spin quantization axis which appears explicitly in the Dirac-Lanczos equation $(12)$.} The correct interpretation of $\Psi_+$ and $\Psi_-$ as an \emph{isospin} doublet  has  been  given for the first time by G\"ursey \cite{GURSE1958-}. This interpretation leads in a simple way to the charge independent theory of strong interactions \cite{GURSE1958-} and to the standard model of electroweak interactions \cite{GSPON1994-}.

    Second, by making a decomposition into a scalar and a vector rather than a decomposition into two singular quaternions, equation $(10)$ describes spin 0 and 1 particles of either positive or negative space-reversal parity \cite{GURSE1950-}.  This led Lanczos to discover the correct equation of a massive spin 1 particle eight years before Proca \cite{LANCZ1929-,GSPON1994-}.

    Third, if the mass $m$ is zero, one gets neutrinos of both handness, as well as Maxwell's equations.\footnote{Strictly speaking, Maxwells's  equations are obtained in the limit where $mA \rightarrow 0$ while $mB$ is kept constant in $(10)$.} 

Finally, if the biquaternions $A$ and $B$ are neither singular or reduced to either a scalar of a vector, Lanczos's equation describes a four-component field which could be associated to spin $\Th$ particles.\footnote{In this paper, we will qualify pairs of four-component biquaternion fields $\{A,B\}$ for which $A\CON{B} \not= 0$  as \emph{general}, in particular to distinguish them from the pairs $\{L,R\}$ for which $L\CON{R}=0$ and which by $(8-9)$ correspond to spin $\Oh$ solutions of Lanczos's equation.}

Therefore, if Lanczos's equation is taken as a fundamental equation of particle physics, the field theory deriving from it describes elementary particles of spin 0, $\Oh$, 1, and potentially $\Th$, which all appear on the same footing within a common framework.  If interactions are allowed according to all possible gauge transformations, the resulting theory contains all the basic ingredients of the current ``Standard model.''  In particular, Lanczos's equation gives a reason why truly elementary fermions come in doublets, and with further assumptions a possible explanation for their replication in several generations \cite{GSPON1998-}.

\section{Rarita-Schwinger formalism in biquaternions}

Since the first attempts by Dirac \cite{DIRAC1936-}, and the more formal approaches of Wigner \cite{WIGNE1939-}, Fierz and Pauli \cite{FIERZ1939-}, and Bargmann \cite{BARGM1948-}, all the proposed theories of high-spin particles are based on the idea that they should generalize the concepts that work in the cases of the Dirac and Proca fields by simply adding more dimensions to the ``configuration space'' that is used to describe the fields associated with the particles. The problem is that beyond spin $\Oh$ the number of the components of the fields grows more rapidly than the spin degree of freedom --- 16 for the Dirac-Rarita-Schwinger field to describe spin $\Th$ when $4+4=8$ are enough to define a four-component spinor and its first derivative.  For this reason it is necessary to subject the fields to ``constraints,'' namely to conditions that hold at a given time.

In this section we will reconsider the Dirac-Rarita-Schwinger theory under the assumption that its formulation with biquaternions may solve, or at least shed some light on the problems that are associated with it.  The main reasons for thinking that this could help are \emph{(i)} that biquaternions are more constraining than the usual matrix formalism because the algebra $\mathbb{B} \sim \cl_{3,0}$ is 8-dimensional while the Dirac algebra $\mathbb{D} \sim\cl_{4,1}$ is 32-dimensional over $\mathbb{R}$, (ii) that the Dirac-Lanczos bispinor $\Psi$ does not transform under the same representation of the Lorentz group as the Dirac bispinor $\psi$ (see, footnote 11), and \emph{(iii)} that the Dirac-Lanczos equation $(12)$ is making explicit the fermionic complex structure that is inherent to spin $\Oh$ particles \cite{GSPON2002-}, and that this complex structure may play a role in a spin $\Th$ formulation based on $(12)$ rather than on the standard Dirac equation.  A first step in this direction was made by Morita, but only for of a vanishing external field, in which case the ``Velo-Zwanziger phenomenon'' and other problems do not arise \cite{MORIT1985-}.

The practical interest of the Dirac-Rarita-Schwinger theory is that, in contrast to the abstract spinor formulation of Fierz and Pauli \cite{FIERZ1939-}, Rarita and Schwinger \cite{RARIT1941-} took the approach of using the spin $\Oh$ Dirac-equation formalism to obtain an explicit formulation that can be generalized to arbitrary-high spin.  For spin $\Th$ the basic idea is to take four distinct Dirac fields $\Psi_\mu$ and to suppose that these fields transform in a Lorentz transformation according to the law
$$
         \Psi_\mu' =  a^\lambda_\mu \mathcal{L} \Psi_\lambda    \eqno(13)
$$
where the $a^\lambda_\mu \in \mathbb{R}$ are the matrix elements of the Lorentz four-vector transformation associated with the tensor index $\mu=0,1,2,3$, and $\mathcal{L}$ the spin $\Oh$ transformation operator of a Dirac bispinor.\footnote{The construction $(13)$ corresponds to the direct product of the 4-component representation $\mathcal{D}(\Oh,\Oh)$ with two copies of the 2-component representation $\mathcal{D}(\Oh,0)$ of the Lorentz group, which yields two copies of the direct sum of the 6-component representation $\mathcal{D}(1,\Oh)$ and of the 2-component representation $\mathcal{D}(0,\Oh)$. Under the rotation subgroup the $SL(2,\mathbb{C})$ representation $\mathcal{D}(0,\Oh)$ remains irreducible and gives the $SU(2)$ representation $\mathcal{D}_{1/2}$, while the $SL(2,\mathbb{C})$ representation $\mathcal{D}(1,\Oh)$ decomposes into the direct sum $\mathcal{D}_{3/2} \oplus \mathcal{D}_{1/2}$.}
 Therefore, it is customarily said that the Rarita-Schwinger field transforms as a ``(four-)vector-(bi)spinor.'' Since this field consists of $4 \times 4=16$ components, it is necessary to impose 8 supplementary conditions to reduce them down to 8.

  In order to transcribe the Rarita-Schwinger equations in quaternions using the Dirac-Lanczos equation, we first rewrite $(12)$ in a form that is better suited to that task.  We therefore introduce the operator
$$
        \CON{\Pi}\A = \CON{\nabla}\Q i\nu - e \CON{\varphi}\Q   \eqno(14)
$$
where $\varphi = \varphi_0 -i \vec{\varphi}$ is the external electromagnetic field.  The minimally coupled \emph{Dirac-Lanczos equation} for a four-component Dirac-Lanczos bispinor $\Psi$ is then \cite{GSPON1998-}
$$
     \CON{\Pi}\Psi = m\Psi^*   ,                                  \eqno(15)
$$
and its conserved probability current density is
$$
C = \Psi^+ \Psi~~,\hskip 1in \SCA \CON\nabla C \LAR=0  ~~.        \eqno(16)
$$

We also introduce a tensor notation for the components of $\nabla$, $\varphi$, and for the four bireal quaternion units
$$
\partial_\mu \in\{ \tfrac{\partial}{\partial x_0},
                  -\tfrac{\partial}{\partial x_n} \}    ~~,  \hskip 0.5in 
       \varphi_\mu  \in \{\varphi_0 , -\varphi_n  \}  ~~,  \hskip 0.5in
                 \epsilon_\mu \in \{1 , ie_n  \}   ~~,                \eqno(17)
$$
where $n=1,2,3$.\footnote{The bireal units, i.e., such that $\epsilon_\mu^+ = \epsilon_\mu$, play in the quaternion formalism a role similar to the $\gamma_\mu$ matrices in the Dirac formalism.}

The Rarita-Schwinger equations are then\footnote{Our tensor conventions are that raising or lowering an index changes the sign of the three spatial components.  Therefore, if $x_\mu \in \{x_0,x_n\}$ then $x^\mu x_\mu = \sum x^\mu x_\mu = x_0^2 -x_n^2$; if  $\partial_\mu \in \{\partial{x_0}, -\partial{x_n} \}$ then $\partial^\mu x_\mu = 4$; if $\epsilon_\mu \in \{1 , ie_n  \}$ then $\CON{\epsilon}^\mu\epsilon_\mu=4$; etc.}
$$
           \CON{\Pi}(\Psi_\mu) = m\Psi_\mu^*          \eqno(18')
$$
$$
           \CON{\epsilon}^\mu\Psi_\mu = 0               \eqno(18'')
$$ 
$$
           \pi^\mu(\Psi_\mu) = 0                       \eqno(18''')
$$
where 
$$
      \pi_\mu\A    = \partial_\mu\Q\nu - e \varphi_\mu             \eqno(19)
$$
are the ``four-vector''-components of $\CON{\Pi}\A$. Equations $(18'')$ and $(18''')$ are the constraints (or supplementary conditions) which reduce the 16 independent components of $(18')$ down to 8.\footnote{The algebraic constraint $(18'')$ is suppressing the $\mathcal{D}(0,\Oh)$ parts of the direct sums $\mathcal{D}(1,\Oh) \oplus \mathcal{D}(0,\Oh)$, while the differential constraint $(18''')$ is expressing the components of the $\mathcal{D}_{1/2}$ part of both $\mathcal{D}(1,\Oh)$ in terms of those of $\mathcal{D}_{3/2}$.}  Since each of the four equations in $(18')$ is a Dirac-Lanczos equations of the type $(15)$, the conserved Rarita-Schwinger probability current density is simply
$$
       C_{RS} = \sum_\mu \Psi_\mu^+ \Psi_\mu~~,    \hskip 1in
                \SCA\CON\nabla C_{RS}\LAR=0~~.                    \eqno(20)
$$

The system $(18)$ seems therefore to have all the necessary properties to provide a suitable description of a spin $\Th$ particle.  However, that this is not the case can be shown by a very simple argument:  Apply the operator $\Pi\A - m\A^*$ (which by $(15)$ gives zero on a Dirac-Lanczos field) to equation $(18'')$ and use the commutator\footnote{The symbol $\odot$ is used to indicate the composition product of two operators.}
$$
              \bigl[  \pi^\mu\A , \CON\Pi\A \bigr] =
     \pi^\mu\A \odot \CON\Pi\A -\CON\Pi\A \odot \pi^\mu\A =
             e \widetilde\Phi(\CON\epsilon^\mu)\Q i\nu             \eqno(21)
$$
where
$$
        \widetilde\Phi\A =\Oh\bigl((\CON\nabla\varphi)\Q 
               + \Q(\varphi\CON\nabla)\bigr)                     \eqno(22)
$$     
is the dual of the electromagnetic field tensor, to conclude that
$$
       \widetilde\Phi(\CON\epsilon^\mu) \Psi_\mu i\nu  = 0 ~~.       \eqno(23)
$$
Therefore, when $\varphi \not= 0$, equation $(23)$ imposes \emph{additional} constraints. This means that the coupling to the external electromagnetic field has reduced the number of the spin states so that the system of equations $(18,23)$ does not provide a good description of a spin $\Th$ particle anymore.

To solve this problem, Fierz and Pauli \cite{FIERZ1939-}, followed by Rarita and Schwinger \cite{RARIT1941-}, proposed to make the coupling to the electromagnetic field in the Lagrangian rather than in the field equations $(18)$.  The procedure for doing this is however not unique.  Nevertheless, we will try to present it in a form that is as general as possible, and then compare our results to those of Velo and Zwanziger~\cite{VELO-1969A}.

What is needed is a system of four single field equations such as $(18')$, but which contain all the constraints and reduce to the system $(18)$ when the external field is zero.  The coupling to the external electromagnetic field is then done in this equation, a procedure that is equivalent to doing it in the corresponding Lagrangian. In the Dirac-Lanczos formalism, a possible Rarita-Schwinger field equation is
$$
  \Bigl[ \Bigl( \CON\Pi\A - m\A^* \Bigr) \delta_\mu^\lambda
 - g \Bigl( \CON\epsilon_\mu \pi^\lambda\A + \pi_\mu\A \CON\epsilon^\lambda \Bigr)
 + g \CON\epsilon_\mu 
        \Bigl( \CON\Pi^*\A + m\A^* \Bigr)
     \CON\epsilon^\lambda \Bigr] \Psi_\lambda  =  0   ~~.     \eqno(24)
$$
Here, $\delta_\mu^\lambda$ is the Kronecker symbol and $g$ is an arbitrary scalar that is $\tfrac{1}{3}$ in \cite{RARIT1941-} and $1$ in \cite{VELO-1969A}.\footnote{Equation $(24)$ is essentially an ansatz, although there could be a canonical way to construct it~\cite{BELLI1972-}.}

We first contract $(24)$ with $\epsilon^\mu$ and use identities such as $\epsilon^\mu\pi_\mu = {\CON\epsilon^*}^\mu\pi_\mu = {\CON\Pi^*}$  and $\SCA \epsilon^\lambda \CON\Pi^*\LAR = \SCA \epsilon^\lambda \epsilon^\mu \pi_\mu\LAR = \pi^\lambda$, as well as $m\A^*\CON\epsilon^\lambda = \epsilon^\lambda m\A^*$ to obtain
$$
  \Bigl[  (4g-1)\epsilon^\lambda m\A^*
   - 2(2g-1)\pi^\lambda\A
   +  (3g-1)\CON\Pi^*\A\CON\epsilon^\lambda \Bigr]\Psi_\lambda = 0 ~~.\eqno(25)
$$
We then contract $(24)$ with the operator $\pi^\mu\A$ and use the identity $\pi^\mu\odot\pi_\mu = \CON\Pi^\mu\odot\CON\Pi_\mu^*$ to get
$$
   \Bigl[ m\Bigl(g\CON\Pi\A\epsilon^\lambda-\pi^\lambda\A\Bigr) \odot \A^*
      + \pi^\lambda\A \odot \CON\Pi\A
      - g \CON\Pi\A\ \odot \pi^\lambda\A \Bigr]\Psi_\lambda = 0 ~~.  \eqno(26)
$$

   When $g=1$ equations $(25)$ and $(26)$ become
$$
  \Bigl[  3\epsilon^\lambda m\A^*
 + 2\Bigl( \CON\Pi^*\A\CON\epsilon^\lambda - \pi^\lambda\A \Bigr)
  \Bigr]\Psi_\lambda = 0                                           \eqno(27)
$$
and
$$
\Bigl[\Bigl(\CON\Pi\A\epsilon^\lambda - \pi^\lambda\A \Bigr) \odot m\A^*
    + \Bigl(\pi^\lambda\A\odot\CON\Pi\A - \CON\Pi\A\odot\pi^\lambda\A \Bigr)
      \Bigr]\Psi_\lambda = 0                                        \eqno(28)
$$
where we see appearing the commutator $(21)$ so that, after complex conjugation,
$$
   m \Bigl(\CON\Pi^*\A\CON\epsilon^\lambda-\pi^\lambda\A\Bigr)\Psi_\lambda
        -   e \widetilde\Phi^*({\CON\epsilon^*}^\lambda)\Psi_\lambda^* i\nu
      = 0  ~~.                                                     \eqno(29)
$$

Equations $(27)$ and $(29)$ are the quaternion equivalent of equations $(2.8)$ and $(2.9)$ of reference \cite{VELO-1969A}.  To proceed, we insert $(29)$ in $(27)$, i.e.,
$$
  \CON\epsilon^\lambda \Psi_\lambda =  \frac{2e}{3m^2}
  \widetilde\Phi({\CON\epsilon}^\lambda)\Psi_\lambda i\nu ~~,       \eqno(30)
$$
which can be used to put  $(29)$ into the form
$$
 \pi^\lambda(\Psi_\lambda) =
  \Bigl(\CON\Pi^*\A + \Th m\A^* \bigr) \frac{2e}{3m^2}
  \widetilde\Phi({\CON\epsilon}^\lambda)\Psi_\lambda i\nu ~~,        \eqno(31)
$$
so that the last two equations enable to rewrite $(24)$ with $g=1$ as
$$
   \Bigl( \CON\Pi\A - m\A^* \Bigr) \Psi_\mu
  =  \Bigl( \pi_\mu\A + \Oh \epsilon_\mu m\A^* \Bigr) \frac{2e}{3m^2}
     \widetilde\Phi({\CON\epsilon}^\lambda)\Psi_\lambda i\nu   ~~.   \eqno(32)
$$
Equations $(30-32)$, which are equivalent to equations $(2.10-2.12)$ of reference \cite{VELO-1969A}, are the final results of this section.  When the electromagnetic field is zero their right hand side vanish and we recover the Rarita-Schwinger equations $(18'-18''')$.  However, contrary to the system $(18)$, the system  $(30-32)$ is consistent in the sense that equations $(30)$ and $(31)$ are now \emph{secondary} constraints that derive from $(32)$, and that there are no unwanted additional constraints such as $(23)$.  Equation $(32)$ is therefore a true equation of motion for a spin $\Th$ particle.

Having derived the Dirac-Rarita-Schwinger equation of motion using the Dirac-Lanczos formalism, we can see by inspecting $(32)$ that its formal structure is identical to that of equation $(2.12)$ of Velo and Zwanziger  \cite{VELO-1969A}, so that the noncausal behavior they discovered will also plague the biquaternion equation $(32)$.

\section{Spin $\Th$ interpretation of the Lanczos's equation}

As we have now confirmed that even when formulated in biquaternions the Rarita-Schwinger equation is not suitable for describing spin $\Th$ particles in an external electromagnetic field, let us try to substantiate the claim that the general four-component quaternion solutions of Lanczos's fundamental equation $(10)$ could correspond to spin $\Th$ particles \cite{GSPON1994-}.

   This can be done by the usual procedure which is to construct the full set of bilinear covariant quantities that are consistent with a field equation and to give their physical interpretation.  To do this in a systematic way we will first recall the bilinear covariants of the spin $\Oh$ interpretation of equation $(10)$, which we rewrite here in the case of a non-vanishing external magnetic field $\varphi$
$$
  \overline\nabla A -e\overline\varphi A = m B  \hskip 1 in
           \nabla B -e         \varphi B = m A ~~,          \eqno(33)
$$
then add the additional bilinear quantities that arise when the spin $\Oh$ restriction to singular solutions is lifted, and finally investigate under which conditions the resulting set would be covariant for particles of spin $\Th$.  

The bilinear covariant quantities which can be built out of $A$ and $B$, assuming that these fields are singular quaternions corresponding to spin $\Oh$ fields transforming under Lorentz transformation as shown in Table 2, are the same as those of the ordinary Dirac theory:

\begin{enumerate}

\item \emph{Polar current}. The four-vector
$$
     C = AA^+  + \CON{BB^+}                           \eqno(34)
$$
is the conserved probability current density. Namely, its divergence
$$
        \SCA \CON{\nabla} C \LAR = 0                  \eqno(35)
$$
is zero even when $\varphi \not= 0$, and its scalar part is the definite positive probability density
$$
     \SCA C \LAR \in \mathbb{R}^+  ~~.                 \eqno(36)
$$
\item \emph{Axial current}. The pseudo four-vector
$$
     \Sigma = AA^+  - \CON{BB^+}                      \eqno(37)
$$
corresponds to the spin density.
\item \emph{Antisymmetric tensor}.  The six-vector
$$
   \vec \pi + i \vec \mu = AB^+ - \CON{AB^+}          \eqno(38)
$$
gives to the electric dipole and magnetic dipole moment densities.
\item \emph{Invariant scalar}.  The complex number $\SCA A^+B \LAR$ corresponds to field combination which enters into the \emph{Lagrangian density} from which Lanczos's equation can be derived,
$$ 
dL = \frac{1}{2} \Bigl\SCA  A^+(\CON{\nabla}-e\CON{\varphi})A - mA^+ B
~~~ ~~~ ~~~ ~~~ ~~~ ~~
$$
$$ 
~~~ ~~~ ~~~ ~~~ ~~~ ~~
                          + B^+(     {\nabla}-e      \varphi )B - mB^+ A
                          + (...)^+              \Bigr\LAR dt  ~~,   \eqno(39)
$$
where $dt$ is the proper time,  as well as to the invariant transition amplitude between ``initial'' and ``final'' states 1 and 2 in perturbation theory
$$
        T_{1,2} = \SCA A_1B_2^+ \LAR \in \mathbb{C}  ~~.          \eqno(40)
$$

\end{enumerate}

In these bilinear covariants, the combinations $A\CON{A}$, $B\CON{B}$, and $A\CON{B}$ do not appear because they are zero as a result of $A$ and $B$ being singular of the form $A=l\sigma$ and $B=r\sigma$ with $l,r \in \mathbb{H}$ since, by $(9)$, they correspond to two-component spinors.  Therefore, if this spin  $\Oh$  restriction is lifted so that $A$ and $B$ form a general pair of biquaternions, there will be additional bilinear covariant quantities.  In fact, it turns out that if one remains within the biquaternion algebra there cannot be other such combinations than $A\CON{A}$, $B\CON{B}$, $A\CON{B}$, and their conjugates.  Using the fact that space-reversal in Lanczos's equation $(33)$ amounts to interchanging $A$ and $B$, we obtain the following additional bilinear quantities:

\begin{enumerate}

\item \emph{Invariant scalar}
$$
              S_P = A\CON{A} + B\CON{B}                     \eqno(41)
$$
\item \emph{Invariant pseudoscalar}
$$
              S_A = A\CON{A} - B\CON{B}                     \eqno(42)
$$
\item \emph{Polar four-vector}
$$
              V_P = A\CON{B}   +  (A\CON{B})^+              \eqno(43)
$$
\item \emph{Axial four-vector}
$$
              V_A = A\CON{B}   -  (A\CON{B})^+              \eqno(44)
$$

\end{enumerate}

To find a possible interpretation to these quantities, we use equation $(33)$ to calculate the divergence of the currents $(43)$ and $(44)$.  It comes
$$
 \SCA \CON\nabla V_P \LAR = 2 \IMA S_P
                           - 2 e \SCA \varphi V_P  \LAR  ~~,    \eqno(45)
$$
and
$$
 \SCA \CON\nabla V_A \LAR = 2 \REA S_P
                           - 2 e \SCA \varphi V_A  \LAR  ~~,   \eqno(46)
$$
which show that the current $V_P$ is conserved when the scalar $S_P$ is real and the electromagnetic field $\varphi=0$.  This suggests that the current $V_P$ can be interpreted as a transition current, implying that (contrary to the spin $\Oh$ solutions) the general four-component solutions of Lanczos's equation are unstable and decay electromagnetically --- which is the case of the spin $\Th$ hadrons.\footnote{The electromagnetic decay fraction of a $\Delta$  into a nucleon is about 0.5 \%. The dominant decay mode (>99 \%) into a nucleon and one pion can be calculated with the charge independent model of strong interactions.}

Therefore, it remains to find out under which conditions the bilinear quantities $(34)$ to $(44)$ are covariant under a representation of the Lorentz group, or at least under one of its subgroups.  Moreover, since we expect the general four-component solutions of Lanczos's equation to correspond to a superposition of four independent rotation eigenstates, they have to be --- at the minimum --- eigenstates of the appropriate spin $\Th$ operator which according to the conventions of Table 1 is
$$
                 J_3\A = \Oh \,i (\nu\Q + 2\Q \nu )  ~~.       \eqno(47)
$$
These conditions are very restrictive, and we know already that we cannot expect them to be satisfied by the spin $\Th$ representation of the Lorentz group given by equation $(6)$.   However, if one does \emph{not} require that the four-component solutions of Lanczos's equation should necessarily transform as the spin $\Th$ representation of the Lorentz group, but accept instead that the action of the Lorentz group on such solutions is what is required to fulfill the stated conditions, there is class of acceptable transformations which have the generic form (see Table 2)
$$
         \mathcal{L}_{3/2}^L\A = \mathcal{L}\Q\mathcal{R}^2  ~~. \eqno(48)
$$
Indeed, such a transformation law is compatible with Lanczos's equation and its bilinear covariants, and it yields the same eigenstates as $\mathcal{L}_{3/2}\A$ given by equation $(6)$ for a rotation around the quantization axis $\nu$. Also, $(48)$ is compatible with the special relativity and quantum scalar products $(1)$ and $(2)$, so that in this respect the behavior of Lanczos's fields of spin 0 to $\Th$ is uniform.  However, the transformations $\mathcal{L}_{3/2}^L\A$ do \emph{not} form a group, except for rotations around an arbitrary quantization axis $\nu$ --- which by Noether's theorem is enough for having a conservation law.

Physically, equation $(48)$ and the corresponding transformation laws in Table~2 mean that it is postulated that while the Lorentz transformations form a group, the action of that group on the solutions of Lanczos's equation is such that its low-spin solutions correspond to the usual spin 0, $\Oh$, and 1 representations of that group, but that this sequence stops at spin $\Th$ for which the action of the Lorentz group is dictated by Lanczos's equation. 

In summary, we have found that the general four-component solutions of Lanczos's fundamental equation $(33)$ could correspond to unstable spin $\Th$ particles.  These solutions are devoid of some of the problems that plague the standard theory based on the Rarita-Schwinger equation because equation $(33)$, or $(10)$, has basically the same structure than Dirac's equation $(7)$.  In particular, since the number of independent complex  components in the $A$ and $B$ fields is $4+4=8$ as in the minimal theory of spin $\Th$ particles, there are no superfluous components which are at the origin of the noncausality of the Rarita-Schwinger theory.  However, these advantages are obtained at the cost of sacrificing the ``Wigner dogma'' that there should be a one-to-one correspondence between unitary representations of the inhomogeneous Lorentz group and elementary particles \cite{WIGNE1939-}.

\begin{table}
\begin{center} 
\begin{tabular}{|c|c|c|c|c|}
\hline
           &                              &                                &           &           \\
{\bf Spin} &  {\bf $A$ field}             &  {\bf $B$ field}               & {\bf N$_A$} & {\bf N$_B$} \\
           &                              &                                &           &           \\
\hline
\hline 
   0    &  $ ~~~ \mathcal{L}\Q\mathcal{L}^+ $ & $                 1\Q             $  & 4 & 2 \\
\hline  
1/2$_+$ &  $     \mathcal{L}\Q\sigma        $ & $     \mathcal{L}^*\Q\sigma       $  & 4 & 4 \\
\hline
1/2$_-$ &  $     \mathcal{L}\Q\CON{\sigma} $ & $     \mathcal{L}^*\Q\CON{\sigma}$  & 4 & 4 \\
\hline
   1    &  $ ~~~ \mathcal{L}\Q\mathcal{L}^+ $ & $ ~~~ \mathcal{L}^*\Q\mathcal{L}^+$  & 4 & 6 \\
\hline
3/2$^L$ &  $ ~~~ \mathcal{L}\Q\mathcal{R}^2 $ & $ ~~~ \mathcal{L}^*\Q\mathcal{R}^2$  & 8 & 8 \\
\hline

\end{tabular}
\end{center}
\caption{\emph{Action of a Lorentz transformation on Lanczos's $A$ and $B$ fields, and numbers N$_A$, N$_B$, of real components in these fields. The Lorentz transformation is characterized by $\mathcal{L}=\mathcal{B}\mathcal{R} \in SL(2,\mathbb{C})$ where $\mathcal{R}=\mathcal{R}^*$ is a spatial rotation and $\mathcal{B}=\mathcal{B}^+$ a Lorentz boost.  The action on the spin $0$, $\Oh$, and $1$ fields corresponds to the respective representations of the Lorentz group. But the action on the four-component solutions of Lanczos's equation, labeled spin }3/2$^L$\emph{, does not correspond to the spin $\Th$ representation of the Lorentz group.}}  
\end{table}

\section{Conclusion: First steps on the ``tenfold way'' ?}

What we have advocated in this paper is a kind of ``Copernician revolution'' consisting of putting the emphasis on a mathematical structure which allows a limited number of spin representations to be described within a common mathematical framework, and on the experimental fact that the infinite series of truly elementary particles of arbitrary-high intrinsic spin that are predicted by well established theories are not observed in nature.

In a way, we come back to Murray Gell-Mann's seminal idea of the ``eightfold way,'' which told the physicist of the 1960s to focus on the baryon and meson octet symmetries rather than on their elusive constituents \cite{GELLM1964-}.  What we suggest today is that the baryon decuplet, which consists of the only spin $\Th$ elementary particles that have been observed, and their exited states, may have a lot tell us --- and that they may give us important clues for a fundamental understanding of the nature of elementary particles.
\footnote{A number of interesting results have recently been found along this path.  For example, in order to explain the systematic clustering of $N$ and $\Delta$ resonances, a new framework for massive spin 3/2 particles has been proposed in which a spin measurement for an unpolarized ensemble of such particles would yield the results $\Th$ with probability one half, and $\Oh$ with probability one half \cite{KIRCH2002-}.  Similarly, in the context of the nucleon spin crisis, a recent proposal is that the correct spin content of the $\Delta$, and the $\Delta$-nucleon mass difference, can be predicted using a spin-dependent quark interaction that reproduces the spin content of the proton \cite{STROB2002-}.  In both examples a departure from the standard formalism is suggested in order to improve the agreement between theory and experiment.}

Concerning the specific idea of this paper --- that the general four-component solutions of Lanczos's fundamental equation may correspond to observed spin $\Th$ particles --- many verifications have to be made beyond those sketched in Section~5.  Moreover, since we do not expect that Lanczos's equation alone is sufficient to give a realistic description of elementary particles, the full interpretation of its spin $\Th$ solutions can only be done in a wider context \cite{GSPON1998-}.  Nevertheless, we can already expect that the symmetry implied by equation $(48)$ may lead to an explanation of the Regge behavior and other features that are characteristic of hadrons, a possibility that we will discuss in a future publication.

\section{Appendix:\\ ~\\ \emph{Quaternion definitions and notations}}

Quaternions were discovered in 1943 by Hamilton who was seeking a generalization to higher dimensions of complex numbers interpreted as doublets of scalars.  While this is not possible for triplets, he found that it is for quadruplets of either real or complex numbers, which yield the field of quaternions $\mathbb{H}$, and the algebra of biquaternions $\mathbb{B}$.  Writing two such quadruplets $A$ and $B$ as scalar-vector doublets $[a;\vec{a}]$ and $[b;\vec{b}]$, the quaternion algebra is obtained by requiring their product to  be associative, and the division to be feasible always, except possibly in some singular cases.  Using contemporary vector notations, this product has the following explicit form
$$
  AB = [a; \vec{a}] [b; \vec{b}] = 
   [ ~ \vec{a}\vec{b} + p ~ \vec{a}\cdot\vec{b}~ ;
     ~ a\vec{b}  +  \vec{a}b + q ~ \vec{a}\times\vec{b} ~ ]  ~~~. \eqno(A.1)
$$
The two constants $p$ and $q$ are related by the equation
$$
                  q^2  + p^3  = 0                        \eqno(A.2)
$$
which shows that there is some residual arbitrariness when defining the product of two quadruplets.\footnote{See, W.R.~Hamilton, Elements of Quaternions, 1866.  Second edition 1899-1901 enlarged by C.J.~Joly.  Reprinted in 1969 by Chelsea Publishing, New York. Vol.I, p.162.} For instance,  taking $p=+1$, $q$ may  be $+i$ or $-i$. This corresponds to the so-called ``Pauli algebra.'' On the other hand, for $p=-1$,  $q$  can  be equal to either $+1$ or $-1$. This is Hamilton's choice, which  is  mathematically more consistent because the imaginary conjugate of a product is equal to the product of the imaginary conjugate of the factors.

The  arbitrariness in the sign of $q$ is due to the non-commutativity  of the quaternion product.  Indeed, changing the order of the factors $A$ and $B$ is equivalent to changing the sign of $q$.  The involution associated with the changing of this sign is called \emph{order reversal} (or simply \emph{reversal}) and is designated by the symbol $\A^\REV$.  When biquaternions are used to represent 
physical quantities in space-time, since $q$ is the sign associated 
with  the  vector product, there is a close  connection  between order reversal and space inversion.

There are three basic quaternion linear automorphisms: \emph{quaternion conjugation} $\CON{\A}$, which reverses the sign of the vector part, \emph{imaginary  conjugation} $\A^*$,  which replaces the  scalar and vector parts by their imaginary conjugate, and \emph{biconjugation} $\A^+$ which is their combination:
$$
        Q \rightarrow \CON{Q}~  = [ s;  -\vec{v}]    ~~~, ~~~
              Q \rightarrow Q^* = [ s^*; \vec{v^*}]  ~~~, ~~~
              Q \rightarrow Q^+ = [ s^*; -\vec{v^*}] ~~~.  \eqno(A.3)
$$

     When  operating  on  a  quaternion  expression,   quaternion 
conjugation reverses the order of the factors.  Thus
$$
              (AB)^* = A^*B^*            \text{~~~~while~~~~}
            \CON{AB} = \CON{B} ~ \CON{A} \text{~~~~and~~~~}
              (AB)^+ = B^+ A^+             ~~~. \eqno(A.4) 
$$

Angle brackets  $\SCA ~ \LAR$ mean that one takes the scalar part of a quaternion expression; round parentheses in the notation $F{\A}$ indicates that $F$ is a function whose argument is conceived to occupy the place marked by $\A$; square brackets $\Q$ are conceived to mark the position to be occupied by a quaternion within a quaternion monomial, e.g., $AB \Q CD$.

For instance, a general spin $\Oh$ Lorentz transformation is
$$
\mathcal{L}_{1/2}\A = \mathcal{L}\Q    ~~~,     \eqno(A.5)
$$
where $\mathcal{L}=\mathcal{B}\mathcal{R} \in SL(2,\mathbb{C})$, i.e., $\CON{\mathcal{R}}\mathcal{R}=\mathcal{B}^*\mathcal{B}=1$, so that $\mathcal{R}\Q$ corresponds to a spinor-rotation and $\mathcal{B}\Q$ to a spinor-boost.

Similarly, the general spin 1 Lorentz transformation for a four-vector is
$$
\mathcal{L}_{1}\A = \mathcal{L}\Q\mathcal{L}^+ = \mathcal{B}\mathcal{R}\Q\CON{\mathcal{R}}\mathcal{B}   ~~~,  \eqno(A.6)
$$
and for a six-vector
$$
\mathcal{L}_{1}'\A = \mathcal{L}^*\Q\mathcal{L}^+ =  \mathcal{B}^*\mathcal{R}\Q\CON{\mathcal{R}}\mathcal{B}  ~~~.  \eqno(A.7)
$$

The power of quaternions stems from their ability to compound between one and eight real numbers which belong to a single (or a few related) tensor quantity(ies) so that many formulas written in biquaternions are simpler than their standard vector, matrix, or tensor counterparts. In general, they enable to dispense of at least one level of tensor indices, and quite often to reduce a few indices tensor to a single entity.  For example, \emph{Proca's equation} for a massive spin 1 field is
$$
                  \CON{\nabla} \vect A  =    B    ~~~,         \eqno(A.8')
$$
$$
     \tfrac{1}{2} (\nabla B  + B^\REV \nabla) = m^2  A    ~~~,     \eqno(A.8'')
$$
where $\nabla = \partial_{it}+\partial_{\vec x}$ is the quaternion four-gradient operator and $A$ the potential which are both assumed to transform as four-vectors according to $(A.6)$, $B$ is a complex vector field transforming as a six-vector according to $(A.7)$, and the symbol $\vect$ means that one takes the vector part of the quaternion product, i.e.,  $\CON{\nabla} \vect A = \CON{\nabla}A - \SCA\CON{\nabla}A\LAR$. Equation $(A.8'')$ shows that the quaternion form of the Proca field \emph{antisymmetric tensor} is the linear function $F\A = \tfrac{1}{2} (\Q B + B^\REV\Q)$,  which should not be confused with the Proca field \emph{bivector} $B$ defined by $(A.8')$, or its reverse  $B^\REV = A \vect \CON{\nabla}$.

For more details on definitions, notations, and elementary applications, see reference \cite{GSPON2001-}.


\begin{thebibliography}{999}

\setlength{\parskip}{0.5mm}

\bibitem{WIGNE1939-} E. Wigner, \emph{On unitary representations of the inhomogeneous Lorentz group}, Ann. of Math. {\bf 40} (1939) 149--204.

\bibitem{BARGM1948-} V. Bargmann and E.P. Wigner, \emph{Group theoretical discussion of relativistic wave equations}, Proc. Nat. Acad. Sci. {\bf 34} (1948) 211-223.

\bibitem{DIRAC1936-} P.A.M. Dirac, \emph{Relativistic wave equations}, Proc. Roy. Soc. {\bf A 155} (1936) 446--459.

\bibitem{FIERZ1939-} M. Fierz and W. Pauli, \emph{On relativistic wave equations for particles of arbitrary spin in an electromagnetic field}, Proc. Roy. Soc. {\bf A 173} (1939) 211--232.

\bibitem{JOHNS1961-} K. Johnson and E.C.G. Sudarshan, \emph{Inconsistency of the local field theory of charged spin 3/2 particles}, Annals of Physics {\bf 13} (1961) 126--145.

\bibitem{VELO-1969A} G. Velo and D. Zwanziger, \emph{Propagation and quantization of Rarita-Schwinger waves in an external potential}, Phys. Rev. {\bf 186} (1969) 1337--1341.

\bibitem{FERRA1992-} A. Ferrara, M. Porrati, and V.L. Telegdi, \emph{$g=2$ as the natural value of the tree-level gyromagnetic ratio of elementary particles}, Phys. Rev. {\bf D 46} (1992) 3529--3537.

\bibitem{VELO-1969B} G. Velo and D. Zwanziger, \emph{Noncausality and other defects of interaction Lagrangians for particles with spin one and higher}, Phys. Rev. {\bf 188} (1969) 2218--2222.

\bibitem{VELO-1978-} G. Velo and A.S. Wightman, eds., Invariant Wave Equations, Lecture Notes in Physics No. 73 (Springer Verlag, New York, 1978) 416 pp.

\bibitem{NOVAE1992-} S.F. Novaes and D. Spehler, \emph{Weyl - van der Waerden spinor techniques for spin~$\Th$ fermions}, Nucl. Phys. {\bf B 371} (1992) 618--636. 

\bibitem{MORIT1985-} K. Morita, \emph{Quaternionic variational formalism for Poincaré gauge theory and supergravity}, Prog. Theor. Phys. {\bf 73} (1985) 999--1015.

\bibitem{PENRO1992-} R. Penrose, \emph{Twistors as spin 3/2 charges}, in A. Zichichi, N. de Sabbata, and N. Sanchez, eds., Gravitation and Modern Cosmology --- P.G. Bergmann's 75th birthday volume (Plenum Press, New York, 1992).

\bibitem{LANCZ1929-} C. Lanczos, \emph{Die  tensoranalytischen  Beziehungen  der Diracschen Gleichung},  Z. f. Phys. {\bf 57}  (1929) 447--473, 474--483, 
484--493. Reprinted and translated {\bf in} W.R. Davis et al., eds., Cornelius Lanczos Collected Published Papers With Commentaries (North Ca\-rolina State University, Raleigh, 1998) Vol. III pages 2-1133 to 2-1225.


\bibitem{GSPON1994-} A. Gsponer and J.-P. Hurni, \emph{Lanczos's equation to replace Dirac's equation~?}, Proc. Int. Cornelius Lanczos Conf., Raleigh, NC, USA (SIAM Publ., 1994) 509--512.  There are a number of typographical errors in this paper. Please ask the authors for a corrected version. 

\bibitem{GSPON1998-} A. Gsponer and J.-P. Hurni, \emph{Lanczos-Einstein-Petiau: From Dirac's equation to non-linear wave mechanics}, {\bf in} W.R. Davis et al., eds., Cornelius Lanczos Collected Published Papers With Commentaries (North Ca\-rolina State University, Raleigh, 1998) Vol. III pages 2-1248 to 2-1277.

\bibitem{RARIT1941-} W. Rarita and J. Schwinger, \emph{On a theory of particles with half-integral spin}, Phys. Rev. {\bf 60} (1941) 61.

\bibitem{GSPON2001-} A. Gsponer and J.-P. Hurni, \emph{Comment on formulating and generalizing Dirac's, Proca's, and Maxwell's equations with biquaternions or Clifford numbers}, Found. Phys. Lett. {\bf 14} (2001) 77--85. For more details, see A. Gsponer and J.-P. Hurni, \emph{The physical heritage of Sir W.R. Hamilton}, presented at the conference: \emph{The Mathematical Heritage of  Sir William Rowan Hamilton} commemorating the sesquicentennial of the  invention of quaternions, Trinity College, Dublin, 17th -- 20th August, 1993, available at http://arXiv.org/abs/math-ph/0201058

\bibitem{BELIN1953-} F.J. Belinfante, \emph{Intrinsic magnetic moment of elementary particles of spin~$\Th$}, Phys. Rev. {\bf 92} (1953) 997--1001. 

\bibitem{GURSE1950-}  F. Gürsey,  \emph{Applications  of  Quaternions  to  Field Equations.}  PhD thesis. (University of London, 1950). 204 pp.

\bibitem{GURSE1958-}  F. Gürsey,  \emph{Relation of charge independence and  baryon conservation to Pauli's transformation}.  Nuovo Cim.   {\bf 7} (1958) 411--415.

\bibitem{GSPON2002-} A. Gsponer, \emph{On the ``equivalence'' of the Maxwell and Dirac equations}, Int. J. Theor. Phys. {\bf 41} (2002) 689--694.

\bibitem{BELLI1972-} J. Bellissard and R. Seiler, \emph{On the Fierz-Pauli equation for particles with spin $\Th$}, Lett. Nuovo Cim. {\bf 5} (1972) 221--225.

\bibitem{GELLM1964-} M. Gell-Mann and Y. Ne'eman, The Eightfold Way (Benjamin, New York, 1964) 327 pp.

\bibitem{KIRCH2002-} M. Kirchbach and D.V. Ahluwalia, \emph{Spacetime structure of massive gravitino}, Phys. Lett. {\bf B 529} (2002) 124--131.

\bibitem{STROB2002-} G.L. Strobel, \emph{Spin dependent quark forces and the spin content of the nucleon}, Int. J. Th. Phys. {\bf 41} (2002) 903--909.


\end{thebibliography}
\end{document}